\documentclass[11pt]{article}
\usepackage{amsmath,amsthm,amsfonts, comment}
\allowdisplaybreaks
\numberwithin{equation}{section}

\theoremstyle{plain}

\newcommand\USU{Department of Mathematics and Statistics,\\ Utah State University,\\ Logan, UT 84322-3900, USA}
\newcommand\UW{Department of Applied Mathematics,\\ University of Washington,\\ Seattle, WA 98195-3925, USA}
\begin{document}

\title{\bf The interaction of long and short waves in dispersive media}

\author{Bernard Deconinck$^\dagger$, Nghiem V. Nguyen$^*$, and Benjamin L. Segal$^\dagger$\\
~\\
$^\dagger$\UW\\
~\\
$^*$\USU}

\date{\today}
\maketitle

\begin{abstract}
The KdV equation models the propagation of long waves in dispersive media, while the NLS equation models the dynamics of narrow-bandwidth wave packets consisting of short dispersive waves. A system that couples the two equations to model the interaction of long and short waves seems attractive and such a system has been studied over the last decades. We evaluate the validity of this system, discussing two main problems. First,
only the system coupling the \textit{linear} Schr\"odinger equation with KdV has been derived in the literature. Second, the time variables appearing in the equations are of a \textit{different} order. It appears that in the manuscripts that study the coupled NLS-KdV system, an assumption has been made that the coupled system can be derived, justifying its mathematical study. In fact, this is true even for the papers where the asymptotic derivation with the problems described above is presented. In addition to discussing these inconsistencies, we present some alternative systems describing the interaction of long and short waves.
\end{abstract}

\section{Introduction}

\renewcommand{\theequation}{\arabic{section}.\arabic{equation}}
\setcounter{section}{1} \setcounter{equation}{0}

Considerable attention ({\em e.g.} \cite{albert, pava1, pava2, chen, corcho, dias}) has been devoted to the following system, which has become known as the cubic nonlinear Schr\"odinger-Korteweg-deVries (NLS-KdV) system:

\begin{equation}
\left\{
\begin{matrix}
\begin{split}
i u_t + u_{xx} + a|u|^2u &= -buv,\\
v_t + c vv_x + v_{xxx} &= -\frac{b}{2}(|u|^2)_x,
\end{split}
\end{matrix}
\right.
\label{NLS-KdV}
\end{equation}
where $x,t \in \mathbb R$, $v(x,t)$ is a real-valued function, $u(x,t)$ is a complex-valued function, and $a$, $b$ and $c$ are real constants. In the context of water waves, the NLS-KdV system was originally introduced by Kawahara \textit{et al.} \cite{KSK} in the form
\begin{equation}
\left\{
\begin{matrix}
\begin{split}
i\bigg(\frac{\partial u}{\partial t_2} + k \frac{\partial u}{\partial x_2}\bigg) + p \frac{\partial^2 u}{\partial x_1^2} &= q uv,\\
\frac{\partial v}{\partial t_3} + \frac{\partial v}{\partial x_3} + \frac{3}{2} v \frac{\partial v}{\partial x_1} + r \frac{\partial^3 v}{\partial x_1^3} &= -s \frac{\partial |u|^2}{\partial x_1},
\end{split}
\end{matrix}
\right.
\label{1975}
\end{equation}
where $k$, $p$, $q$, $r$ and $s$ are real constants, $x_n=\epsilon^n x$, and $t_n = \epsilon^n t$. Here $\epsilon$ is the small parameter in terms of which the asymptotic expansions were performed.
This system couples two of the most studied equations in mathematical physics: the KdV equation describes the unidirectional propagation of long, nonlinear dispersive waves, while the cubic nonlinear Schr\"odinger equation governs the slowly varying modulation of a narrow bandwidth train of short waves. Both equations are completely integrable \cite{ggkm, sz}. As such, the system (\ref{1975}) is interesting both from a mathematical and a physical point of view.

However, there are several concerns regarding the above system which have been ignored thus far.  Even though many authors (the papers \cite{albert, pava1, pava2, chen, corcho, dias} are but a small sample of the relevant literature) have studied different mathematical aspects of system (\ref{NLS-KdV}), there exists a tendency to cross reference without checking the details of the original derivation.  Tracing through a plethora of references, the exact derivation of system (\ref{NLS-KdV}) is nowhere to be found. We were led eventually to the paper by Kawahara \textit{et al.} \cite{KSK} which appears to be where the system (\ref{1975}) was first introduced in the context of water waves.  Notice that the first equation in (\ref{1975}) is \textit{linear} whilst that in (\ref{NLS-KdV}) is \textit{nonlinear}. Further, the time scales appearing in (\ref{1975}) are inconsistent, with the dynamics of the second equation of (\ref{1975}) appearing on a slower time scale than that of the first equation. More on this is discussed below. The same is true for the derivation in the context of plasma physics, see \cite{appert, ikezi, nishikawa}, where references lead back to \cite{Zak} and the system (\ref{NLS-KdV}) is not found in any form. Thus it appears that works heretofore studying (\ref{NLS-KdV}) are investigating the mathematical aspects of a \textit{hypothetical} system that has not been derived consistently. Of course, these mathematical considerations are perfectly valid in their own right, but it should be stated that to this point, the results presented are yet to be shown relevant in the context of any application.

Even the derivation of system (\ref{1975}) in \cite{KSK} is problematic.  Starting from the Euler water wave problem, the authors introduce multiple spatial and temporal scales $x_n = \epsilon^n x$ and $t_n= \epsilon^n t$ with $x_0=x$ and $t_0=t$ to expand the velocity potential and surface elevation functions in an asymptotic series, while assuming that the waves travel in one direction.  At the orders $\epsilon^4$ and $\epsilon^5$, the equations of (\ref{1975}) arise as a consequence of eliminating secular terms. It is immediately clear that the system (\ref{1975}) is troublesome as the two equations appear at different time and spatial scales.  This is dealt with in \cite{KSK} by rewriting the final equations in terms of the first-order slow variables $x_1$ and $t_1$: $t_2=\epsilon t_1$, $t_3=\epsilon^2 t_1$ and $x_2=\epsilon x_1$, $x_3=\epsilon^2 x_1$. Of course this is an inconsistent argument: the different equations encountered to this point are obtained by equating terms at the same order of $\epsilon$. Reintroducing $\epsilon$ at a later point invalidates all calculations to this point.  
The main problems with the applicability of (\ref{NLS-KdV}) can be summarized as thus:

\begin{itemize}

\item[(A)] Only a system coupling the \textit{linear} Schr\"odinger (LS) equation with the KdV equation has ever been derived in the form (\ref{1975}) (see also \cite{funakoshi, ikezi, nishikawa}).

\item[(B)] In the two coupled equations, two \textit{different} time scales appear.

\end{itemize}

Many authors refer to one or multiple of \cite{funakoshi, ikezi, KSK, nishikawa} and each other to motivate the use of the system (\ref{NLS-KdV}), while apparently the details and the results presented in  these papers are ignored.

In addition to analyzing (A) and (B) in more detail, we propose some alternative systems to (\ref{1975}).  Our starting point for the asymptotic expansions is the fifth-order KdV equation as introduced below in (\ref{fullequation}). It would seem more natural to start from the Euler water wave equations or the equations describing waves in plasmas (and this is done in a forthcoming paper), but this offers no direct advantage: the fifth-order KdV equation incorporates all the physical effects one wants to consider in a derivation of the coupled system: one-dimensional propagation, dispersion, nonlinearity, the possibility of second-harmonic resonance, {\em etc}. It is well known \cite{kdv2nls} that even the classical KdV equation may be used as the starting point to derive the NLS equation. Expanding solutions of the fifth-order KdV equation as a power series in $\epsilon$, we write the coefficient of each power of $\epsilon$ as a superposition of a long and a short wave, as is done in \cite{KSK}. Next, secular contributions are eliminated at each order, resulting in a set of two equations at each order.
At order $\epsilon^4$, one of the equations is the LS equation, {\em i.e.}, the first equation appearing in (\ref{1975}). At order $\epsilon^5$, one of the equations is the KdV equation, the second one in (\ref{1975}). This is exactly what the authors in \cite{KSK} obtained when they combined the two equations that appear at \textit{different} orders of the asymptotic expansion. To some extent, this confirms our claim that the fifth-order KdV equation is a suitable laboratory for the investigation at hand.  Our point now is clear: instead of using one equation from each of the orders $\epsilon^4$ and $\epsilon^5$, we will use two equations that appear at the \textit{same} order of the expansion.


Our calculations and the results obtained from them indicate the impossibility for the derivation of (\ref{NLS-KdV}) in the context of any physical system describing the interaction of long and short waves in dispersive media. It appears impossible even to derive (\ref{1975}) with both equations appearing at the same order.  It is important to note that in \cite{DS2}, by working with the full Euler equations in three spatial dimensions, {\em i.e.}, $(x,y,z,t) \in\mathbb R^3\times \mathbb R$, the authors obtain the ODE version of (\ref{1975}). That is, using a specific traveling-wave solution ansatz in the Euler equations, and after expanding the solution in an asymptotic series, the authors obtain a system of ODEs. This is the same system of ODEs one finds using a traveling-wave ansatz directly on (\ref{1975}). This is expected, of course, as the traveling-wave ansatz effectively eliminates the inconsistent time derivatives in (\ref{1975}). This does not authenticate the derivation of the PDE system (\ref{1975}) where the two equations appear \textit{at the same order}.

\section{Summary of results}
\renewcommand{\theequation}{\arabic{section}.\arabic{equation}}
\setcounter{section}{2} \setcounter{equation}{0}

Although the vocabulary used below is that from the theory of surface water waves, our considerations are equally valid in the context of plasma physics. We do not include here the full set of Euler equations governing the surface water wave problem or the system for plasma waves consisting of the fluid equations coupled with Maxwell's equations. The interested reader can find both systems in \cite{infeldrowlands}, for instance.

The validity of these systems is undisputed, but so is their complexity. Because of this, simpler asymptotic models that focus on the incorporation of less than the full gamut of physical effects are frequently used. One-dimensional and unidirectional waves are frequently observed and studied, both in the long- and short-wave regime. The case of long waves in shallow water, for instance, gives rise to the celebrated Korteweg-de Vries (KdV) equation
\begin{equation}\label{kdv}
u_t -\lambda u_x -3 uu_x + (\tau-1/3) u_{xxx} =0,
\end{equation}
where $\tau= \kappa/g h^2$ is a dimensionless measure of the importance of surface tension {\em vs.} the effect of gravity. Here $g$ is the acceleration of gravity, $h$ is the depth of the undisturbed water surface, $\kappa$ is the coefficient of surface tension, and $\lambda$ is a real number associated with the Froude number \cite{as}.

Another, more complicated equation incorporating more physical effects, was derived by Johnson in 2002 \cite{J}. This equation reads
\begin{equation}
u_t + \lambda u_x + c_0uu_x + \alpha u^2u_x + \beta u_xu_{xx} +\gamma u u_{xxx} + c_1 u_{xxx} + c_2 u_{xxxxx} =0,
\label{fullequation}
\end{equation}
with the seven constants $\lambda$, $c_0$, $\alpha$, $\beta$, $\gamma$, $c_1$ and $c_2$  specified in \cite{J}. With different values of the constants, this equation may also be derived when $\tau$ in (\ref{kdv}) is near $1/3$, at which point the derivation leading to (\ref{kdv}) breaks down \cite{sunetal}. Equation (\ref{fullequation}) is the starting point of our calculations.

We expand the solution $u(x,t)$ of (\ref{fullequation}) asymptotically in the form
\begin{equation*}
u(x,t) = \sum_{j=1}^{\infty} \epsilon^j u_j(x,t).
\end{equation*}
Introducing the stretched variables
\begin{equation}
\xi= \epsilon(x-c_g t);  \ \ \ \ \ \tau_j = \epsilon^j t,
\label{changeofvariables}
\end{equation}
and keeping terms of orders up to $\epsilon^5$ in (\ref{fullequation}) we obtain the following expression:

\begin{equation}\nonumber
\begin{split}
&\epsilon u_{1t} + \epsilon^2 u_{2t} + \epsilon^3 u_{3t} + \epsilon^4 u_{4t} + \epsilon^5 u_{5t} - \epsilon^2 c_g u_{1\xi} - \epsilon^3 c_g u_{2\xi} - \epsilon^4 c_g u_{3\xi}
- \epsilon^5 c_g u_{4\xi} + \epsilon^2 u_{1\tau_1}\\
&+ \epsilon^3 u_{2\tau_1} + \epsilon^4 u_{3\tau_1} + \epsilon^5 u_{4\tau_1} + \epsilon^3 u_{1\tau_2} + \epsilon^4 u_{2\tau_2} + \epsilon^5 u_{3\tau_2} + \epsilon^4 u_{1\tau_3} + \epsilon^5 u_{2\tau_3} + \epsilon^5 u_{1\tau_4}= \\
&-\epsilon\big(\lambda u_{1x} +c_1u_{1xxx} + c_2 u_{1xxxxx}\big)
- \epsilon^2 \big(\lambda u_{2x} + \lambda u_{1\xi}+ c_2 u_{2xxxxx} + 5c_2 u_{1xxxx\xi}+\\
&+\beta u_{1x} u_{1xx}+ \gamma u_1 u_{1xxx} +3c_1 u_{1xx\xi}
+ c_1 u_{2xxx}+ c_0 u_1 u_{1x} \big)\\
& -\epsilon^3 \big(\lambda u_{3x} +\lambda u_{2\xi} + c_2 u_{3xxxxx} + 5c_2 u_{2xxxx\xi} + 10 c_2 u_{1xxx\xi\xi}+ \alpha u_1^2 u_{1x}+ \beta u_{1x} u_{2xx}\\
& + 2\beta u_{1x} u_{1x\xi} + \beta u_{2x} u_{1xx} +\beta u_{1\xi} u_{1xx} + \gamma u_1 u_{2xx} + 3\gamma u_1 u_{1xx\xi} + \gamma u_2 u_{1xxx} + c_0 u_1 u_{2x}\\
& + c_0 u_1 u_{1\xi}+ c_0 u_2 u_{1x} + c_1 u_{3xxx}+ 3 c_1 u_{2xx\xi} + 3c_1 u_{1x\xi\xi}\big)\\
& -\epsilon^4\big(\lambda u_{4x} + \lambda u_{3\xi} + c_2 u_{4xxxxx} + 5 c_2 u_{3xxxx\xi} + 10c_2 u_{2xxx\xi\xi} + 10c_2 u_{1xx\xi\xi\xi} + \alpha u_1^2 u_{2x} \\
&+\alpha u_1^2 u_{1\xi}+2\alpha u_1 u_2 u_{1x} + \beta u_{1x} u_{3xx} + 2\beta u_{1x} u_{2x\xi} + \beta u_{1x}u_{1\xi\xi} + \beta u_{2x} u_{2xx} + 2\beta u_{2x} u_{1x\xi}\\
&+ \beta u_{3x} u_{1xx}+\beta u_{1\xi} u_{2xx} + 2\beta u_{1\xi} u_{1x\xi} + \beta u_{2\xi} u_{1xx} + \gamma u_1 u_{3xxx} + 3\gamma u_1 u_{2xx\xi} + 3\gamma u_1 u_{1x\xi\xi}\\
&+ \gamma u_2 u_{2xxx} + 3\gamma u_2 u_{1xx\xi} +\gamma u_3 u_{1xxx} + c_0 u_1 u_{3x} + c_0 u_1 u_{2\xi} + c_0 u_2 u_{2x} + c_0 u_2 u_{1\xi}\\
& + c_0 u_3 u_{1x} + c_1 u_{4xxx} + 3c_1 u_{3xx\xi} + 3c_1 u_{2x\xi\xi} + c_1 u_{1\xi\xi\xi}\big)
\end{split}
\end{equation}

\begin{equation}
\begin{split}
& -\epsilon^5 \big(\lambda u_{5x}+ \lambda u_{4\xi} + c_2u_{5xxxxx} + 5c_2 u_{4xxxx\xi} + 10c_2 u_{3xxx\xi\xi} + 10c_2 u_{2xx\xi\xi\xi} +5c_2 u_{1x\xi\xi\xi\xi} \\
&+ c_2 u_{1\xi\xi\xi\xi\xi} +\alpha u_1^2u_{3x} + \alpha u_1^2 u_{2\xi} +\alpha u_2^2 u_{1x} + 2\alpha u_1 u_2 u_{2x} + 2\alpha u_1 u_2 u_{1\xi} + 2\alpha u_1 u_3 u_{1x}\\
&+\beta u_{1x} u_{4xx} + 2\beta u_{1x}u_{3x\xi} + \beta u_{1x}u_{2\xi\xi} + \beta u_{2x} u_{3xx} + 2\beta u_{2x}u_{2x\xi} + \beta u_{2x}u_{1\xi\xi} + \beta u_{3x}u_{2xx}\\
&+ 2\beta u_{3x} u_{1x\xi} + \beta u_{4x}u_{1xx} + \beta u_{1\xi} u_{3xx} + 2\beta u_{1\xi}u_{2x\xi} + \beta u_{1\xi} u_{1\xi\xi} +\beta u_{2\xi}u_{2xx} + 2\beta u_{2\xi}u_{1x\xi}\\
&+ \beta u_{3\xi} u_{1xx} + \gamma u_1 u_{4xxx}+ 3\gamma u_1 u_{3xx\xi} + 3\gamma u_1 u_{2x\xi\xi} + \gamma u_1 u_{1\xi\xi\xi} + \gamma u_2 u_{3xxx} + 3\gamma u_2 u_{2xx\xi} \\
&+3\gamma u_2 u_{1x\xi\xi} +\gamma u_3 u_{2xxx} + 3\gamma u_3 u_{1xx\xi} + \gamma u_4 u_{1xxx}+ c_0u_1 u_{4x} + c_0u_1 u_{3x} + c_0u_2 u_{3x}\\
& +c_0 u_2 u_{2\xi} + c_0u_3 u_{2x} + c_0u_3 u_{1\xi} + c_0 u_4 u_{1x} + c_1 u_{5xxx} + 3c_1u_{4xx\xi} + 3c_1u_{3x\xi\xi} + c_1 u_{2\xi\xi\xi}\big).
\end{split}
\end{equation}

\sloppypar \noindent As in \cite{funakoshi, KSK}, we let $u_1(x,t)$ be the linear superposition of a long wave $C_1(\xi,\tau_1,\tau_2,\tau_3)$ and a short, narrow-bandwidth wave $e^{ikx -i\omega t}A(\xi,\tau_1,\tau_2,\tau_3) + e^{-ikx +i\omega t}\overline A(\xi,\tau_1,\tau_2,\tau_3)$:
\begin{equation}\label{trad}
u_1(x,t)=C_1(\xi,\tau_1,\tau_2,\tau_3)+e^{ikx -i\omega t}A(\xi,\tau_1,\tau_2,\tau_3) + e^{-ikx +i\omega t}\overline A(\xi,\tau_1,\tau_2,\tau_3),
\end{equation}
with $k\neq 0$. Next, we equate different powers of $\epsilon$ to zero. Requiring the absence of secular terms results in additional constraints.  A few remarks are in order.

\vspace*{0.1in}

\noindent {\bf Remarks.}

\begin{enumerate}

\item To obtain a KdV-type equation, the presence of the term $u_{j\xi\xi\xi}$ is required. This term cannot occur at order lower than $\epsilon^4$ (with $j=1$).  Thus, the inclusion of the real-valued function
$C_j(\xi,\tau_1,\tau_2,\tau_3)$ is necessary in the expression for $u_j(x,t)$, and it is necessary to proceed to order $\epsilon^4$ at least, in order to find a KdV-type equation.

\item In order to obtain an NLS-type equation, one needs the term $|A|^2 A$. The lowest order at which this term can be found is $\epsilon^3$, from the term $\alpha u_1^2 u_{1x}$.  It may appear that such nonlinearities can be achieved at higher orders too. For instance, at order $\epsilon^4$, the terms $\alpha u_1^2 u_{2x}$ and $2\alpha u_1u_2 u_{1x }$ can potentially yield a contribution containing $|A|^2 A$ if  $u_2$ also contains $e^{ikx -i\omega t}A(\xi,\tau_1,\tau_2,\tau_3) + e^{-ikx +i\omega t}\overline A(\xi,\tau_1,\tau_2,\tau_3)$. This, however, implies the presence of $A_{\xi\xi\xi}$, which is inconsistent with the NLS equation.

\end{enumerate}

The summary of both remarks is that the NLS-KdV system (\ref{NLS-KdV}) cannot be derived with the traditional ansatz (\ref{trad}) used in \cite{funakoshi, ikezi,KSK,nishikawa}, starting from a generic system which has nonlinearities that are quadratic, cubic, {\em etc.}

\vspace*{0.1in}

In what follows, we derive the system
\begin{equation}
\left\{
\begin{matrix}
A_{\tau_3} +   (c_1 -10c_2 k^2) A_{\xi\xi\xi} - \beta k^2 AC_{\xi} - 3\gamma k^2 A_{\xi}C + c_0 (AC)_{\xi}=0,\\
C_{\tau_3} +c_0CC_{\xi} + c_1 C_{\xi\xi\xi} =-(\beta k^2 +c_0-3\gamma k^2)(|A|^2)_{\xi}.\\
\end{matrix}
\right.
\label{system0}
\end{equation}
This is arguably the simplest system that is consistent with the constraints described in the above remarks, where the second equation is the KdV equation. If $\beta\neq 3\gamma$, it appears that this system is not Hamiltonian and does not have any conserved quantities. For $\beta=3\gamma$, the system may be rewritten as

\begin{equation}
\left\{
\begin{matrix}
u_t + 2bu_x + au_{xxx}= -2 b(uv)_x,\\
v_t + bv_x + bvv_x + c v_{xxx} = -b(|u|^2)_x,
\end{matrix}
\right.
\label{system1}
\end{equation}

\noindent by using a simple change of variables and renaming the constants. Note that $v(x,t)$ is a real-valued function, while $u(x,t)$ is complex valued. The system (\ref{system1}) is Hamiltonian:
\begin{equation}
\label{ham1}
\frac{\partial}{\partial t}\left(
\begin{array}{c}
  u \\
  v
\end{array}
\right)
= J_1 \left(
\begin{array}{c}
 \delta H_3/\delta u \\
 \delta H_3/\delta v
\end{array}
\right), \ \ \ \ \ \mbox{with} \ \ \ \ \ J_1\left(
\begin{array}{c}
  u \\
  v
\end{array}
\right) =\left(
\begin{array}{c}
  u_x \\
  v_x
\end{array}
\right),
\end{equation}
and has at least four conserved quantities:
\begin{equation*}
H_0(u) = \int_{-\infty}^{\infty} u\, dx, \ \ \ \ \ H_1(v) = \int_{-\infty}^{\infty} v\, dx,\ \ \ \ \
H_2(u,v) = \int_{-\infty}^{\infty} \big(|u|^2 + v^2\big)\,dx,
\end{equation*}
\begin{equation*}
H_3(u,v) = \int_{-\infty}^{\infty} \left(\frac{a}{2}|u_x|^2 + \frac{c}{2}v^2_x- \frac{b}{6} v^3 -b |u|^2 v - b|u|^2 -\frac{b}{2} v^2\right)\,dx.
\end{equation*}
The first two, $H_0(u)$ and $H_1(v)$, are Casimirs of the system. It should be noted that (\ref{fullequation}) is not Hamiltonian unless $\beta=2\gamma$. In general (\ref{fullequation}) has only the one conserved quantity $\int_{-\infty}^\infty u\,dx$. If $\beta=2\gamma$, the equation is Hamiltonian and has three conserved quantities. The Hamiltonian structure is only found for (\ref{system0}) if a {\em different} constraint on the parameters is satisfied. Since our perturbation procedure outlined below is not a Hamiltonian one (nor can it be since (\ref{system0}) is not Hamiltonian in general), we are not guaranteed to maintain any Hamiltonian structure even for $\beta=2\gamma$. The presence of the extra structure for $\beta=3\gamma$ in (\ref{system1}) is a bonus. It is of interest to realize that if the constants in (\ref{fullequation}) are related back to the water wave problem, the constraint $\beta=3\gamma$ can only be satisfied for a specific (non-zero) value of the coefficient of surface tension. It would be of interest to use the Hamiltonian structure of (\ref{fullequation}) with $\beta=2\gamma$ as the starting point for a Hamiltonian perturbation method (see {\em e.g.} \cite{craig1, craig2}) to derive a Hamiltonian system coupling the interaction of long and short waves. This is not pursued here.


Next, using the relations at order $\epsilon^3$ to alter the system at order $\epsilon^5$, we obtain a BBM-like system \cite{bbm} describing the interaction of long and short waves (for $\beta=3\gamma$):
\begin{equation}
\left\{
\begin{matrix}
u_t + 2b u_x - \mu u_{xxt} = -2b(uv)_x,\\
v_t + bv_x + bvv_x - \sigma v_{xxt} = -b(|u|^2)_x.
\end{matrix}
\right.
\label{system2}
\end{equation}
As for (\ref{system1}), $u(x,t)$ is complex valued while $v(x,t)$ takes on real values only.

The system (\ref{system2}) has at least four conserved quantities
\begin{equation*}
\mathcal{H}_0(u) = \int_{-\infty}^{\infty} u\,dx, \ \ \ \ \ \mathcal{H}_1(v) = \int_{-\infty}^{\infty} v \,dx
, \ \ \ \ \
\mathcal{H}_2(u,v) = \int_{-\infty}^{\infty} \left(|u|^2 + \mu |u_x|^2 + v^2+ \sigma v_x^2\right)dx,
\end{equation*}
\begin{equation*}
\mathcal{H}_3(u,v) = \int_{-\infty}^{\infty} \left(b|u|^2 + \frac{b}{2}v^2 + \frac{b}{6} v^3 +b|u|^2 v\right)dx,
\end{equation*}
and is Hamiltonian:
\begin{equation}
\label{ham2}
\frac{\partial}{\partial t}\left(
\begin{array}{c}
  u \\
  v
\end{array}
\right)
= J_2 \left(
\begin{array}{c}
 \delta H_3/\delta u \\
 \delta H_3/\delta v
\end{array}
\right), \ \ \ \ \ \mbox{with} \ \ \ \ \ J_2\left(
\begin{array}{c}
  u \\
  v
\end{array}
\right) =-\left(
\begin{array}{c}
  (1-\mu \partial_x^2)^{-1}u_x \\
  (1-\sigma \partial_x^2)^{-1}v_x
\end{array}
\right).
\end{equation}

All results about the systems (\ref{system1}) and (\ref{system2}) stated here are derived in the next two sections.

\section{Derivation of the systems (\ref{system1}) and (\ref{system2})}
\renewcommand{\theequation}{\arabic{section}.\arabic{equation}}
\setcounter{section}{3} \setcounter{equation}{0}

The reader can verify that the term $u_1(x,t)$ in the perturbation series for $u(x,t)$ has to vanish in order for all necessary effects to be incorporated.  Thus we equate $u_1\equiv 0$ and we start the expansion at order $\epsilon^2$.

At {\bf second order}, we find
\begin{equation*}
u_{2t} + \lambda u_{2x} + c_1u_{2xxx} + c_2 u_{2xxxxx}=0.
\end{equation*}

\noindent With $u_2(x,t)$ given by
$$
u_2(x,t) = e^{ikx-i\omega t} A(\xi,\tau_1,\tau_2,\tau_3) + c.c. + C(\xi,\tau_1,\tau_2,\tau_3),
$$
(where $c.c.$ stands for complex conjugate), we find the dispersion relation
\begin{equation}
\omega(k) = \lambda k -c_1 k^3 + c_2 k^5.
\label{w(k)}
\end{equation}

At {\bf third order}, we obtain
\begin{equation*}
\begin{split}
&u_{3t} + \lambda u_{3x} + c_1u_{3xxx} + c_2 u_{3xxxxx}= c_g u_{2\xi}-\lambda u_{2\xi} - u_{2\tau_1} - 5 c_2 u_{2xxxx\xi} - 3c_1 u_{2xx\xi}\\
&= (c_g -\lambda) C_{\xi} - C_{\tau_1} + e^{ikx -i\omega t}\left((c_g -\lambda) A_{\xi} - A_{\tau_1} - 5c_2 k^4 A_{\xi} + 3 c_1 k^2 A_{\xi}\right)+c.c.
\end{split}
\end{equation*}

\noindent In order for the solution $u(x,t)$ to be bounded, we impose the secularity conditions

\begin{equation}
(c_g-\lambda) C_{\xi} - C_{\tau_1}=0,
\label{2}
\end{equation}
\begin{equation}
A_{\tau_1} - \big[(c_g-\lambda) + 5c_2 k^4 - 3 c_1 k^2\big]A_{\xi}=0.
\label{3}
\end{equation}

\noindent This leaves us
\begin{equation*}
u_{3t} + \lambda u_{3x} + c_1u_{3xxx} + c_2 u_{3xxxxx}=0.
\end{equation*}
The third-order solution $u_3(x,t)$ is a superposition of expressions of the form $e^{ikx-iwt} B(\xi,\tau_1,\tau_2,\tau_3) + c.c. + D(\xi,\tau_1,\tau_2,\tau_3)$ where the functions $B$ and $D$ will be
determined at the next order. Since we aim to derive the simplest system coupling long and short waves, we choose  $u_3(x,t)\equiv 0$.

At {\bf fourth order} in $\epsilon$, we get
\begin{equation}
\begin{split}
&u_{4t} + \lambda u_{4x} + c_1u_{4xxx} + c_2 u_{4xxxxx}\\
&=-u_{2\tau_2} -10 c_2 u_{2xxx\xi\xi} -\beta u_{2x}u_{2xx} -\gamma u_2 u_{2xxx} -c_0 u_2 u_{2x}-3c_1u_{2x\xi\xi} \\
&= -C_{\tau_2} + e^{ikx -i\omega t}[ - A_{\tau_2} + ik(10 c_2k^2 - 3c_1) A_{\xi\xi} + ik (\gamma k^2 -c_0) AC]\\
&+ e^{2ikx -2i\omega t} A^2(i\beta k^3 +i\gamma k^3 - ic_0k)+c.c.
\label{sweet}
\end{split}
\end{equation}

\noindent Once again we impose secularity conditions:

\begin{equation}
 C_{\tau_2}=0,
\label{4}
\end{equation}
\begin{equation}
i A_{\tau_2} +   k(10c_2 k^2 - 3c_1) A_{\xi\xi} + k(\gamma k^2 -c_0) AC =0.
\label{5}
\end{equation}

\noindent With these conditions imposed, the method of undetermined coefficients applied to (\ref{sweet}) gives
$$
u_4(x,t) = \frac{\beta k^2 + \gamma k^2 -c_0}{6k^2(5c_2k^2 -c_1)} e^{2ikx -2i\omega t} A^2+c.c.,
$$
provided there is no resonance:
\begin{equation}
 5c_2k^2 - c_1 \not= 0.
 \label{resonance}
\end{equation}

At {\bf fifth order},
\begin{equation*}
\begin{split}
&u_{5t}+\lambda u_{5x} + c_1u_{5xxx} + c_2 u_{5xxxxx}\\
&=c_g u_{4\xi}-u_{4\tau_1}-\lambda u_{4\xi}-u_{2\tau_3}- 5c_2u_{4xxxx\xi} -10 c_2 u_{2xx\xi\xi\xi}\\
&-2\beta u_{2x}u_{2x\xi}-\beta u_{2\xi}u_{2xx} -3\gamma u_2 u_{2xx\xi} -c_0 u_2 u_{2\xi}-3c_1 u_{4xx\xi} -c_1u_{2\xi\xi\xi} \\
&= -C_{\tau_3} -c_0CC_{\xi} - c_1 C_{\xi\xi\xi} -(\beta k^2 +c_0-3\gamma k^2)(|A|^2)_{\xi}+ H\!H(e^{ikx -i\omega t})\\
& +e^{ikx -i\omega t}[ - A_{\tau_3} + (10 c_2k^2 - c_1) A_{\xi\xi\xi} + \beta k^2 AC_{\xi} + 3\gamma k^2 A_{\xi}C-c_0 (AC)_{\xi}] + c.c.,
\end{split}
\end{equation*}
where $H\!H$ denote higher harmonics: terms that are proportional to higher ({\em i.e.}, $\geq 2$) powers of $\exp (ikx -i\omega t)$. The fifth-order equation leads to the secularity conditions

\begin{equation}
C_{\tau_3} +c_0CC_{\xi} + c_1 C_{\xi\xi\xi} =-(\beta k^2 +c_0-3\gamma k^2)(|A|^2)_{\xi},
\label{6}
\end{equation}
\begin{equation}
A_{\tau_3} +   (c_1 -10c_2 k^2) A_{\xi\xi\xi} - \beta k^2 AC_{\xi} - 3\gamma k^2 A_{\xi}C + c_0 (AC)_{\xi} =0,
\label{7}
\end{equation}
which make up system (\ref{system0}).

Choosing
\begin{equation}
\beta k^2 = 3\gamma k^2=-c_0,
\label{8}
\end{equation}
the two equations (\ref{6}) and (\ref{7}) can be rewritten to give the following system in terms of the slow variables $(\tau_3,\xi)$:
\begin{equation*}
\left\{
\begin{matrix}
\begin{split}
&A_{\tau_3} + (c_1 -10c_2 k^2) A_{\xi\xi\xi}= -2 c_0 (AC)_{\xi},\\
&C_{\tau_3} + c_0CC_{\xi} + c_1 C_{\xi\xi\xi} =-c_0 (|A|^2)_{\xi}.
\end{split}
\end{matrix}
\right.
\end{equation*}
The first equality in (\ref{8}) is a choice, while the second is easily realized using a scaling of the variables in (\ref{fullequation}). As already mentioned above, the choice of the first equality results in a Hamiltonian system, as shown in the next section. A further change of variables $u=A$, $v=C-1$, and a relabeling of $\xi$ as $x$ and $\tau_3$ as t, results in
\begin{equation}
\left\{
\begin{matrix}
u_t + (c_1 -10c_2 k^2) u_{xxx} + 2c_0 u_x = -2 c_0 (uv)_x,\\
v_t + c_0 v_x + c_0vv_x + c_1 v_{xxx} =-c_0 (|u|^2)_x,
\end{matrix}
\right.
\label{9}
\end{equation}
which is the announced system (\ref{system1}). It should be noted that imposing the non-resonance condition is necessary in order to find a Hamiltonian system of equations. If resonance occurs, (\ref{sweet}) results in an extra secularity condition $i\beta k^3+i\gamma k^3-i c_0 k=0$, which contradicts (\ref{8}).

\vspace*{0.1in}

Recall that at order $\epsilon^3$ in the derivation of system (\ref{system1}), we have the relations (\ref{2}) and (\ref{3}). In terms of the new variables $(x,t)$, we can replace the $x-$derivative in (\ref{9}) by
\begin{equation}
\left\{
\begin{matrix}
\begin{split}
&u_x = \frac{- \epsilon^2 u_t}{(\lambda-c_g) - 5c_2 k^4 + 3 c_1 k^2},\\
&v_x =  \frac{- \epsilon^2 v_t}{\lambda-c_g},\\
\end{split}
\end{matrix}
\right.
\end{equation}
provided that $(\lambda-c_g) - 5c_2 k^4 + 3 c_1 k^2\not= 0$ and $ \lambda \not=c_g.$  This leads to the new system
\begin{equation}
\left\{
\begin{matrix}
\displaystyle u_t + 2c_0 u_x -\frac{c_1 -10c_2 k^2}{(\lambda-c_g) - 5c_2 k^4 + 3 c_1 k^2}\epsilon^2 u_{xxt} &= -2 c_0 (uv)_x,\\
\displaystyle v_t + c_0 v_x + c_0vv_x - \frac{c_1}{\lambda -c_g}\epsilon^2 v_{xxt} &=-c_0 (|u|^2)_x,
\end{matrix}
\right.
\label{10}
\end{equation}
which is system (\ref{system2}) after relabeling the constants.

The system (\ref{system2}) above is not derived using consistent asymptotics, as the expansion parameter $\epsilon$ appears in the final equation. In other words, one of the criticisms we leveled at (\ref{NLS-KdV}) and (\ref{1975}) also applies to (\ref{system2}), as it does to the BBM equation. However, as for the BBM equation, the use of inconsistent asymptotics in this case leads to a system with superior well-posedness properties. This is not something that can be said for the systems (\ref{NLS-KdV}) or (\ref{1975}).

\section{Hamiltonian structure and Conservation Laws}
\renewcommand{\theequation}{\arabic{section}.\arabic{equation}}
\setcounter{section}{4} \setcounter{equation}{0}
For this entire section, we assume $u,v \in C_0^\infty (\mathbb R)$.\\

\noindent \textbf{Claim.} \textit{The following functionals are conserved quantities for system (\ref{system1})}.
\begin{equation*}
H_0(u) = \int_{-\infty}^{\infty} u\, dx, \ \ \ \ \ H_1(v) = \int_{-\infty}^{\infty} v\, dx, \ \ \ \ \
H_2(u,v) = \int_{-\infty}^{\infty} \left(|u|^2 + v^2 \right)dx,
\end{equation*}
\begin{equation*}
H_3(u,v) = \int_{-\infty}^{\infty} \left(\frac{a}{2}|u_x|^2 + \frac{c}{2}v^2_x- \frac{b}{6} v^3 -b |u|^2 v - b|u|^2 -\frac{b}{2} v^2\right)dx.
\end{equation*}
\begin{proof}
Integrating both sides of system (\ref{system1}) over the whole real line to obtain
\begin{equation*}
\begin{split}
\int_{-\infty}^{\infty} u_t dx= \frac{d}{dt} H_0(u) &= \int_{-\infty}^{\infty} \left(-a u_{xx} - 2buv - 2b u\right)_x dx =0,\\
\int_{-\infty}^{\infty} v_t dx= \frac{d}{dt} H_1(v) &= \int_{-\infty}^{\infty}\left(- b v -\frac{b}{2} v^2 - cv_{xx} -b|u|^2 \right)_x dx =0.
\end{split}
\end{equation*}
Taking the time derivative of $H_2(u,v)$ we have
\begin{equation*}
\begin{split}
\frac{d}{dt} H_2(u,v) &= \int_{-\infty}^\infty \left(u_t \overline{u}+ u\overline{u}_t + 2 vv_t\right) dx\\
&= \int_{-\infty}^{\infty}\overline{u} (-a u_{xxx} - 2b(uv)_x -2bu_x) dx + \int_{-\infty}^{\infty} u(-a\overline{u}_{xxx} - 2b(\overline{u} v)_x - 2b\overline{u}_x)dx\\
& ~~~+ 2 \int_{-\infty}^{\infty} v\big(-b v_x-b vv_x - c v_{xxx} - b(|u|^2)_x\big) dx\\
&= \int_{-\infty}^{\infty} \left(- 2b v(|u|^2)_x -2b u(\overline{u} v)_x - 2b \overline{u} (uv)_x \right)\,dx=0.\\
\end{split}
\end{equation*}
A similar calculation with $H_3(u,v)$ gives
\begin{equation*}
\begin{split}
\frac{d}{dt} H_3(u,v) &= \int_{-\infty}^\infty\left(\frac{a}{2} u_x \overline{u}_{xt} + \frac{a}{2} u_{xt}\overline{u}_x + c v_x v_{xt} - \frac{b}{2} v^2 v_t -b v_t |u|^2
-b v u_t \overline{u} -bv u \overline{u}_t\right)\,dx\\
&~~~+\int_{-\infty}^\infty\left( -b u\overline{u}_t -bu_t \overline{u} -b vv_t\right)\,dx\\
&=\int_{-\infty}^{\infty} \left( -\frac{a}{2} u_{xx} -b uv -bu\big) \big( -a\overline{u}_{xxx} - 2b(\overline{u}v)_x -2b \overline{u}_x\right)\,dx\\
& ~~~+ \int_{-\infty}^\infty \left( -\frac{a}{2} \overline{u}_{xx}
 -b\overline{u}v -b\overline{u}\big) \big( -au_{xxx} - 2b(uv)_x -2b u_x\right)\,dx\\
& ~~~+ \int_{-\infty}^{\infty} \left(cv_{xx} +\frac{b}{2} v^2 +b|u|^2+bv ) \big( cv_{xxx} +\frac{b}{2} (v^2)_x +b(|u|^2)_x + bv_x \right)\,dx\\
&=\frac{1}{2}\int_{-\infty}^{\infty} \left( |au_{xx} + 2b uv + 2bu|^2 + \left(cv_{xx} + \frac{b}{2} v^2 +b|u|^2 + bv\right)^2\right)_x dx =0.
\end{split}
\end{equation*}
As the time derivatives of $H_0$, $H_1$, $H_2$ and $H_3$ are zero, they must be constant in time. This concludes the proof.
\end{proof}

The next claim is shown in the same straightforward way.
\vspace*{0.1in}

\noindent \textbf{Claim:} \textit{The following functionals are conserved quantities for system (\ref{system2})}.
\begin{equation*}
\mathcal{H}_0(u) = \int_{-\infty}^{\infty} u\,dx, \ \ \ \mathcal{H}_1(v) = \int_{-\infty}^{\infty} v\,dx,
\ \ \
\mathcal{H}_2(u,v) = \int_{-\infty}^{\infty} \left( \mu |u_x|^2 +\sigma v^2_x +|u|^2 + v^2\right)dx,
\end{equation*}
\begin{equation*}
\mathcal{H}_3(u,v) = \int_{-\infty}^{\infty} \left(b|u|^2 + \frac{b}{2}v^2 + \frac{b}{6} v^3 +b|u|^2 v\right)dx.
\end{equation*}

\noindent Note that the multiplicative constant $b$ can be omitted from $\mathcal{H}_3(u,v)$, but it is necessary for $\mathcal{H}_3(u,v)$ to be the Hamiltonian for (\ref{system2}).

\begin{proof}
Integrating both sides of (\ref{system2}) over the whole real line we obtain
\begin{equation*}
\begin{split}
\int_{-\infty}^{\infty} u_t dx= \frac{d}{dt} \mathcal{H}_0(u) &= \int_{-\infty}^{\infty} \left(\mu u_{xxt} -2bu_x - 2b(uv)_x\right) dx =0,\\
\int_{-\infty}^{\infty} v_t dx= \frac{d}{dt} \mathcal{H}_1(v) &= \int_{-\infty}^{\infty}\left( -bv_x -\frac{b}{2} (v^2)_x + \sigma v_{xxt} -b(|u|^2)_x \right)dx =0.
\end{split}
\end{equation*}
Taking the time derivative of $\mathcal{H}_2$ we have
\begin{equation*}
\begin{split}
&\frac{d}{dt} \mathcal{H}_2(u,v) = \int_{-\infty}^\infty (\mu u_x \overline{u}_{xt} + \mu u_{xt}\overline{u}_x +2\sigma v_{x}v_{xt} + 2v v_t +  u\overline{u}_t +  u_t \overline{u}) dx\\
&= \int_{-\infty}^{\infty}\bigg( -\mu u_{xx}\overline{u}_t - \mu\overline{u}_{xx} u_t - 2 \sigma v_{xx} v_t + 2v(-bv_x -bvv_x + \sigma v_{xxt} - b(|u|^2)_x\big)\bigg)dx\\
&~~+ \int_{-\infty}^{\infty}\bigg( u\big(-2b \overline{u}_x +\mu\overline{u}_{xxt} -2b(\overline{u}v)_x\big) +\overline{u}\big(-2b u_x +\mu u_{xxt} - 2b(uv)_x\big) \bigg)dx\\
& =\int_{-\infty}^{\infty}\bigg( -2b v (|u|^2)_x -2b u(\overline{u}v)_x -2b \overline{u}(uv)_x \bigg)dx =0.
\end{split}
\end{equation*}
Checking the conservation of $\mathcal{H}_3(u,v)$ requires a bit more work. Define
$$
z(x,t) =\int^x_{-\infty} u_t(y,t) dy \ \ \ \ \ \mbox{and} \ \ \ \ \ w(x,t) = \int^x_{-\infty} v_t(y,t)dy.
$$
Then $z_x= u_t$ and $w_x = v_t.$  Multiplying the second equation in (\ref{system2}) by $w$ and integrating over the whole real line, we obtain
$$
\int_{-\infty}^{\infty}\left(ww_x + w\left(bv+\frac{b}{2} v^2\right)_x +\sigma w w_{xxx} + b w(|u|^2)_x\right)dx=0,
$$
which gives
\begin{equation}
\int_{-\infty}^{\infty}\left(\left(\frac{b}{2} v^2 +\frac{b}{6}v^3\right)_t + bv_t |u|^2 \right) dx =0.
\label{11}
\end{equation}
Likewise, multiplying the first equation in (\ref{system2}) by $\overline{z}$ and the conjugate of it by $z$, summing and integrating the result over the whole real line, we get
\begin{equation}
\int^\infty_{\infty}\bigg( b(|u|^2)_t + b v(|u|^2)_t \bigg)dx =0.
\label{12}
\end{equation}
Adding (\ref{11}) and (\ref{12}), we arrive at the desired result:
$$
\frac{d}{dt} \mathcal{H}_3(u,v)=0.
$$
As the time derivatives of $\mathcal{H}_0$, $\mathcal{H}_1$, $\mathcal{H}_2$ and $\mathcal{H}_3$ are all zero, they must be constant in time.
\end{proof}

To verify the Hamiltonian structure of the two systems, we calculate the Frechet derivatives of $H_3(u,v)$ and $\mathcal{H}_3(u,v)$.
\begin{equation*}
\begin{split}
&\left(\frac{\delta H_3}{\delta u}, \frac{\delta H_3}{\delta v}\right)= (-a u_{xx} - 2buv -2b u, -c v_{xx} - \frac{b}{2} v^2 - b|u|^2 -bv),\\
&\left(\frac{\delta \mathcal{H}_3}{\delta u}, \frac{\delta \mathcal{H}_3}{\delta v}\right) = ( 2bu + 2buv, bv+\frac{b}{2} v^2 + b|u|^2).
\end{split}
\end{equation*}
The first system (\ref{system1}) is rewritten as
\begin{equation*}
(u_t,v_t) = \left(-a u_{xxx} - 2b(uv)_x - 2b u_x, -c v_{xxx} - \frac{b}{2} (v^2)_x -b(|u|^2)_x -b v_x \right),
\end{equation*}
so that (\ref{system1}) is Hamiltonian with Poisson structure $J_1$ and Hamiltonian $H_3(u,v)$, see (\ref{ham1}). The operator $J_1$ is a well-known Poisson operator, as it is a vectorized version of that for the KdV equation \cite{as}.

Similarly, from the second system (\ref{system2}), we can write
\begin{equation*}
(u_t,v_t) = \bigg(-(1-\mu \partial_x^2)^{-1} (2bu +2buv)_x , -(1-\sigma \partial_x^2)^{-1}(bv +\frac{b}{2} v^2 + b|u|^2)_x \bigg).
\end{equation*}
Thus, (\ref{system2}) is Hamiltonian with Poisson structure $J_2$ and Hamiltonian $\mathcal{H}_3 (u,v)$ as written in (\ref{ham2}). As for $J_1$, $J_2$ is a valid Poisson operator, being a vectorized version of that for the BBM equation \cite{olver}.

\section*{Acknowledgements.} The authors wish to thank Shu-ming Sun (Virginia Tech) for interesting conversations. BD acknowledges support from the National Science Foundation (NSF-DMS-1522677). The work of BLS was supported by a University of Washington Applied Mathematics Boeing Fellowship and an ARCS foundation Seattle Chapter fellowship. Any opinions, findings, and conclusions or recommendations expressed in this material are those of the authors and do not necessarily reflect the views of the funding sources.

\bibliographystyle{siam}

\end{document}